\documentstyle[12pt]{article}
\begin{document}
\title{Application of Statisitcal Physics\\ to\\ Politics}
\author{Serge Galam \\
Laboratoire des Milieux D\'{e}sordonn\'{e}s
et H\'{e}t\'{e}rog\`{e}nes\thanks{Laboratoire associ\'{e} au
CNRS (UMR n$^{\circ}$ 7603)},\\
Tour 13 - Case 86, 4 place Jussieu, \\ 75252 Paris Cedex 05, France\\
(galam@ccr.jussieu.fr}
\date{July 8,1999}
\maketitle
\begin{abstract}
The concept and technics of real space renormalization group are
applied to study majority rule voting in hierarchical structures.
It is found that democratic voting can lead to totalitarianism by
keeping in power a small minority. Conditions of this paradox are
analyzed and singled out. Indeed majority rule produces critical
thresholds to absolute power. Values of these thresholds can vary
from $50\%$ up to at least $77\%$. The associated underlying
mechanism could provide an explanation for both former apparent
eternity of communist leaderships as well as their sudden
collapse.

\end{abstract}
\newpage
\section{Setting the limits}

Modern theory of critical phenomena is based on the fundamental concepts of 
universality and irrelevant variables [1]. These two concepts mean that different 
physical systems, like for instance a magnet and a liquid, behave the same way 
when passing from one macroscopic state to another macroscopic state. Well 
known examples are the magnet becoming a para-magnet, the liquid, a gas, 
and even may be the creation of the universe from nothing with the big bang.
Most of the microscopic properties of the physical compounds involved turn out 
to be irrelevant for describing the macroscopic change which in turn appears 
to be universal. While the number of physical systems undergoing phase
transitions is infinite, all associated phase transitions can be described in 
terms of only a small finite number of universality classes. Only a few 
parameters, like space dimensionality, determine which universality class 
the system belongs to. The abstract and general nature of the statistical physics 
framework makes it tempting to extend such notions to non-physical systems, 
and in particular to social systems, for which, in many  cases, 
there exists an interplay between microscopic properties and macroscopic realities.
 
Nevertheless the two fields of physical sciences and social sciences are rather 
different. However, the process of going in parallel from one atom and one human
to respectively several atoms in bulk and a social group has much in common.
More precisely, it is the hypothesis behind the present approach. It is worth 
stressing that we are not claiming our models will explain all aspects of human 
behavior. Like any modeling effort, it is appropriate only to some classes 
of phenomena in social science and not to others.

However such an approach should be carefully controlled. To just map a physical 
theory built for a physical reality, onto a social reality, could be at best a 
nice metaphore, but without predictability, and at worst a misleading and 
wrong social theory. Physics has been successfull in describing macroscopic 
behavior using properties of the constituant microscopic elements. 
The task here, is to borrow from physics those techniques and concepts 
used to tackle the complexity of aggregations.  The challenge is then 
to build a collective theory of social behavior along similar lines, but 
 within the specific constraints of the psycho-social reality. The constant 
danger is for the theorist to stay in physics, using a social terminology 
and a physical formalism. The contribution from physics should thus be 
restricted to qualitative guidelines for the mathematical modeling of complex 
social realities. Such a limitation does not make the program less ambitious.

\section{Real space: from physics to politics}

In this paper we present an application of statistical physics to political sciences
[2]. We apply the concept and technics of real space renormalization group [1] to study
the majority rule voting process within hierarchical structures. In particular
we focus on the conditions for a given political party to get for sure, full power at
the hierarchy top level.

We find that majority rule voting produces critical threshold to total power.
Having an initial support above the critical threshold guarantees full power at
the top. The value of the critical threshold to power is a function of the voting
structure, namely the size of voting groups and the number of voting
hierarchical levels. Using these results a new explanation is given
to the sudden and abrupt fall of eastern former communist parties.

Here we apply the real space renormalization group
scheme of collective phenomena in physics to a radically
different reality, namely a social one  We emphasis on other 
technical aspects than the ones usually used in physics.
While there it is an abstract and formal tool to
study phase transitions, here we 
associate a political reality to each step of the renormalization
group transformation. On this basis it is worth to stress we apply
statistical physics to political sciences not as a qualitative
metaphor but indeed  as a guide to build a quantitative model to study
the effect of majority rule voting on hte democratic representation of 
groups within a hierarchical organization. 

Like in physics we start from local cells constituted
by a small number of degrees of freedom, here individuals. Similarly
to Ising spins, to keep the analysis simple, people
can choose only between two political
tendencies A and B. Associated proportions of A and B support
in the system, a political group, a firm, a network, a society,
are supposed to be known.
We denote them by
$p_0$ for the overall A-support and  $1-p_0$ for the B-support. We
are assuming each member does have an opinion.

Once formed, each cell elects a representative, either an A or a B
using a majority rule. These elected people (the equivalent of the
super spin rescaled to an Ising one in real space renormalization
group) constitute the first hierarchical level of the organization
called level-1. New cells are then formed at level-1 from these
elected people. They in turn elect new representatives to build
level-2. This process is repeated again and again.

In physics the rescaled degree of freedom is fictius while here it
is a real person. We are not using a theoretical scheme to embody
some complex features but instead we are building up a real
organization where each voting step is real.
Moreover at odd with physics the number of
iterations in the renormalization process are finite and the focus
is on the stable fixed points.

The following of the paper is as follows. In next section we
present and study the case of 3-person cells. A critical threshold
to power is singled out. It equals $50\%$. A minority is found to
self-eliminate within a few voting levels. A making sense bias in
the voting rule is then introduced in Section 3. A one vote bonus
is added to the tendency already in power in cases of A-B
equality. For 4-person cells, this bias shift the critical
threshold to power from $50\%$ to $77\%$. Size effects are
analyzed in Section 4. Analytic formulas are then derived. In
particular, given an initial support $p_0$ for the A, the number
of voting levels necessary to their self-elimination is obtained.
Section 5 puts the results in a more practical perspective.
Last section contains a discussion about  both former apparent
eternity of communist leaderships as well as their sudden
collapse. Some.perspectives are outlined.

\section{The simplest fair case}

We start from a population distributed among two tendencies A and B
with respectively $p_0$ and  $1-p_0$ proportions
within the the system. It could be either a political group, a firm,
or a whole society. At this stage each member does have an opinion.
From now on we will use the political language.

Cells are constituted by randomly aggregating group of 3 persons.
It could be by home localization or working place. Each cell then
elects a representative using majority rule. To have an A elected
requires either 3 A or 2 A in the cell. Otherwise it is a b who is
elected. these elected persons constitute the first level of the
hierarchy denoted level-1. The same process of cell forming can be
repeated within level-1 from the elected persons. Making the cell
to vote produces an additional level, namely level-2. We can go on
the same way from a level-n to a level-(n+1). The probability to
have an A elected at level (n+1) is then,
\begin{equation}
p_{n+1}\equiv P_3(p_n)=p_n^3+3 p_n^2 (1-p_n) \ ,
\end{equation}
where $p_n$ is the proportion of elected A persons at level-n.

We call $P_3(p_n)$ the voting function. It has 3 fixed points
$p_d=0$, $p_{c,3}=\frac{1}{2}$ and $p_t=1$. First one corresponds
to the disappearance of the A. Last one $p_t$ represents the
totalitarian situation where only A are present. Both are stable.
At contrast $p_c$ is unstable. It determines the threshold to full
power. Starting from $p_0<\frac{1}{2}$ repeating voting leads
towards 0 while the flow is in direction of 1 for
$p_0>\frac{1}{2}$.

Therefore majority rule voting produces the self-elimination of
any proportion of the A-tendency as long as $p_0<\frac{1}{2}$.
However the democratic self-elimination to be completed requires a
sufficient number of voting levels.

At this stage the instrumental question is to determine the number
of levels required to ensure full leadership to the initial larger
tendency. The analysis will turn relevant to reality only if this
level numbers is small. Most organizations has only a few level,
and always less than 10.

For instance starting from $p_0=0.43$ we get successively
$p_1=0.40 $, $p_2=0.35$, $p_3=0.28$, $p_4=0.20$, $p_5=0.10$,
$p_6=0.03$ down to $p_7=0.00$. Therefore 7 levels are sufficient
to self-eliminate $43\%$ of the population.

Tough the aggregating voting process eliminate a tendency it stays
democratic since it is the leading tendency (more than $50\%$ )
which after all gets the total leadership of the organization. It
is worth to notice the symmetry with respect to A nd B tendencies.

\section{The simplest killing case}

In real world things are not as fair as above and often it turns
out very hard, if not impossible, to change an organization
leadership. We will now illustrate this situation.

Considering yet the simplest case we constitute groups of 4
persons instead of 3. The salient new feature is the 2A-2B
configuration for which there exists no clear majority. In most
social situations it is well admitted that to change a policy
required a clear cut majority. In case of no decision, things will
stay as they are. It is a bias in favor of the already existing.
Often this bias is achieved giving for instance, one additional
vote to the committee president.

Along this line, the voting function becomes non symmetrical.
Assuming the B were in power, for an A to be elected at level
$n+1$ we have,
\begin{equation}
p_{n+1}\equiv P_4(p_n)=p_n^4+4 p_n^3 (1-p_n) \ ,
\end{equation}
where $p_n$ is the proportion of elected A persons at level-n. At
contrast for a B it is,
\begin{equation}
1- P_4(p_n)=p_n^4+4 p_n^3 (1-p_n)+2 p_n^2 (1-p_n)^2 \ ,
\end{equation}
where last term embodies the bias in favor of B. From Eqs (2) and (3) the stable
fixed points are still 0 and 1. However the unstable one is drastically shifted to,
\begin{equation}
p_{c,4}=\frac{1+\sqrt{13}}{6} \ ,
\end{equation}
which makes the threshold to power to the A about $77\%$. Moreover
the process of self-elimination is accelerated. For instance from
$p_0=0.69 $ we have the series  $p_1=0.63 $, $p_2=0.53 $,
$p_3=0.36 $, $p_4=0.14 $, $p_5=0.01 $, and $p_6=0.00$. It shows
that using an a priori reasonable bias in favor of the B turns a
majority rule democratic voting to a totalitarian outcome. Indeed
to get to power the A must pass over $77\%$ of support which
almost out of a possibility. Above series shows how $63\%$ of a
population disappears from the leadership within only 5 voting
levels.

\section{Larger voting groups}

Most real organizations work with voting cells larger than 3 or 4.
To account for this size variable we generalize above scheme to
cells with r voting persons. We then have to determine the voting
function $p_{n+1}=P_r(p_{n})$. Using a majority rule it becomes,
\begin{equation}
P_r(p_n)=\sum_{l=r}^{l=m}\frac{r!}{l!(r-l)!} p_n^l(1+p_n)^{r-l}\ ,
\end{equation}
where $m=\frac{r+1}{2}$ for odd r and $m=\frac{r+1}{2}$ for even r which thus
accounts for the B-bias.

The two stable fixed points $p_d=0$ and $p_t=1$ are preserved with enlarging
the group size. However while the unstable one stays $p_{c,r}=\frac{1}{2}$ for odd
sizes, it varies with r for even sizes. It starts at $p_{c,4}=\frac{1+\sqrt{13}}{6}$
with the limit
$p_{c,r}\rightarrow \frac{1}{2}$ when $r \rightarrow \infty$.

We can then calculate analytically the critical number of levels $n_c$ at which
$p_{n_c}=\epsilon$ with
$\epsilon$ being a very small number. It determines the level of confidence
of the prediction to have no A elected.. One way to evaluate $n_c$
is to expand the voting function
$p_n=P_r({n-1})$ around the unstable fixed point $p_{c,r}$,
\begin{equation}
p_n\approx p_{c,r}+(p_{n-1}-p_{c,r}) \lambda_r \ ,
\end{equation}
where $\lambda_r \equiv \frac{dP_r(p_n)}{dp_n}|_{p_{c,r}}$ with
$P_r(p_c)=p_{c,r}$. Rewriting
last equation as,
\begin{equation}
p_n-p_{c,r}\approx (p_{n-1}-p_{c,r}) \lambda_r \ ,
\end{equation}
we can then iterate the process to get,
\begin{equation}
p_n-p_{c,r}\approx (p_0-p_{c,r}) \lambda_r^n \ ,
\end{equation}
from which we get the critical number of levels $n_c$ at which $p_n=\epsilon$.
Taking the $\ln$ on both
side of Eq. (8) gives,
\begin{equation}
n\approx -\frac{\ln (p_c-p_0)}{\ln \lambda_r} +n_0   \ ,
\end{equation}
where $n_0\equiv \frac{\ln (p_{c,r} -\epsilon)}{\ln \lambda_r}$ is only
valid not too far from $p_{c,r}$. However it turns out to be a rather good estimate
even down to the sable fixed point 0 by making it equal to 1 while taking
the integer part of Eq .(9). For a more accurate calculation of $n_0$ see [3].

\section{The magic formula}

Most organizations don't change their structure at every election
or decision event. They are set once and then don't change any
longer. The number of hierarchical levels is thus fixed and
constant. Therefore to make our analysis useful we have to invert
the question  on``how many levels are needed to eliminate a tendency"
onto ``given n levels what is the necessary overall support to get
full power". 

It is worth to keep in mind that situations for respectively  A and B tendencies
are not always symetric. Here we stress the dynamics of voting
with respect to the A. To implement theis operative question, we invert Eq.(7) to,
\begin{equation}
p_0=p_{c,r}+(p_n-p_{c,r}) \lambda_r^{-n} \ .
\end{equation}
indeed two critical thresholds now appears.
First one, the disappearence threshold $p_{d,r}^n$ which gives the value
of support under which the A disappears for sure at the top level leadership. 
It is given by Eq. (10) putting $p_n=0$,
\begin{equation}
p_{d,r}^n=p_{c,r}(1-\lambda_r^{-n}) \ .
\end{equation}
In parallel putting $p_n=1$ give the second threshold $p_{f,r}^n$ above which 
the A get full and total power. Using Eq.(11), we get,
\begin{equation}
p_{f,r}^n=p_{d,r}^n+\lambda_r^{-n}\ .
\end{equation}
which shows the appearence of a new region for $p_{d,r}^n<p_0<p_{f,r}^n$.
In that region the A neither disappear totally nor get full power
($p_n$ is neither $0$ nor $1$). It is
therefore a coexistence region where some democraty is prevailing since results
of the election process are only probabilistic. No tendency is sure of winning
making alternating leadership a reality. However as seen from Eq.(12), this
democratic region shrinks as a power law $\lambda_r^{-n}$ of 
the number n of hierarchical levels.
Having a small number of levels thus  puts higher the threshold to a total 
reversal of power but simultaneously lowers the threshold for non existence.

Again, above formulas are approximates since we have neglected
corrections in the vicinity of the stable fixed points. However
they give the right qualitative behavior. Actually  $p_{d,r}^n$ fits to $n+1$
and  $p_{f,r}^n$ to $n+2$. For more accurate formulas see [3].

To get a practical feeling of what Eqs. (11) and (12) means, let us illustrate them for
the case $r=4$ where we have $\lambda =1.64$ and $p_{c,4}=\frac{1+\sqrt{13}}{6}$.
Considering 3, 4, 5, 6 and 7 level organizations, $p_{d,r}^n$ is equals to 
respectively $0.59$, $0.66$, $0.70$, $0.73$ and $0.74$. In parallel $p_{f,r}^n$
equals $0.82$, $0.80$, $0.79$, $0.78$ and $0.78$.
These series emphasizes drastically the totalitarian character of the voting process.

\section{Some prospective}

Up to now we have treated very simple cases to single out main trends produced 
by democratic voting aggregating over several levels. In particular we have 
singled out
the existence of critical thresholds to full power. Moreover these tresholds
are not necesseraly symetric for both tendencies in competition. In the biased 4-cell
case it is around $0.77\%$ for the A. 

Such asymetries are indeed always present in most realistic situations in which more than
two groups are competing. Let us consider for instance the case of three competing
groups A, B and C. Assuming a 3-cell case, now the ABC configuration is unsolved
using majority rule as it was for the precedent two tendencies AABB 4-cell configuration.
For the AB case we made the bias in favor of the group already in power, like
giving an additional vote to the comitee president. 

For multi-group competitions typically the bias results from parties agreement.
Usually the two largest parties, say A and B are hostile while 
the smallest one C would compromise with either one of them. Then the ABC configuration
gives a C elected. In such a case, we need 2A or 2B to elect respectively an A or a B.
Otherwise a C is elected. Therefore the elective function for A and B are the same as
for the AB 3-cell model. It means that the critical threshold to full power to A and B
is $50\%$. In otherwords for initial A and B supports less than $50\%$ the C get full
power. The required number of levels is obtained from above formulas.

It is possible to generalize to as many groups as wanted. The analysis becomes more 
heavy but the mean features of voting flows towards fixed point are preserved.

\section{Conclusion}

To conclude we comment on some possible new explanation to the recent generalized
auto-collapse of eastern communist parties. Up to this historical and drastic
event, communist parties seemed eternal with the same leadership ever. Once they 
collapsed most explanations were related to both an opportunistic change within the
various organizations together with the end of the soviet army threat.

May be the explanation is different and related to our hierachical model. Indeed
communist organizations are based on the structutral concept
of democratic centralism which nothing else than a tree-like hierachy with a rather high 
critical threshold to power. Suppose it was of order of $80\%$ like in our 4-cell case. 
We could than consider that the internal opposition to the orthodox leadership
did grew a lot and massively over several decades to evantually reach and pass
the critical threshold with the associated sudden outrise of the internal opposition.
Therefore the at once collapse of eatern communist pparties would have been the result
of a very long and solid phenomena. Such an explanation does not oppose to additional
constraints but emphazie on the internal mechanism within these organizations.

At this stage it is of importance to stress that modeling social and political
phenomena is not stating some absolute truth but insted to single out
some basic trend within very complexe situations.


\end{document}